\documentclass[preprint,aps,prd,groupedaddress,showpacs,floatfix,%
nofootinbib]{revtex4}
\usepackage{dcolumn}% Align table columns on decimal point
\usepackage{graphicx}
\usepackage{epsfig}
\usepackage{hyperref}
\usepackage{graphicx}% Include figure files
\usepackage{dcolumn}% Align table columns on decimal point
\usepackage{amssymb,eucal}
\usepackage{longtable}

\newcommand\Tr{{\rm Tr }}

\newcommand\tr{{\rm tr }}

\begin{document}
\title{Skyrmion properties from holographic instantons of large size}
\author{Fen Zuo$^{a,b,c}$\footnote{Email: \textsf{fen.zuo@ba.infn.it}},
        Feng-Kun Guo$^d$\footnote{Email: \textsf{fkguo@hiskp.uni-bonn.de}},
        and Tao Huang$^{b,c}$\footnote{Email: \textsf{huangtao@ihep.ac.cn}}} %
\affiliation{
  $^a$Istituto Nazionale di Fisica Nucleare, Sezione di Bari, Italy\\
  $^b$Institute of High Energy Physics, CAS, Beijing 100049, China\\
  $^c$Theoretical Physics Research Center for Science Facilities, CAS, Beijing 100049, China\\
  $^d$Helmholtz-Institut f\"ur Strahlen- und Kernphysik and Bethe Center for
      Theoretical Physics, Universit\"at Bonn, D-53115 Bonn, Germany}
\begin{abstract}
Based on the intanton approximation of Skyrmion and
recent progresses on the holographic approach, we study baryon properties using
Skyrmions generated from holographic instantons. First we employ Atiyah and
Manton's early observation to show that the instanton approximation gives the
correct infrared behavior for the Skyrmion solution, and thus those of the
electromagnetic form factors of the nucleon. %, as opposed to the claims of Cohen et al.
We then use Skyrmions generated from flat-space instanton solutions to study
various baryon properties, treating the instanton size as an arbitrary variable. It
is found that, a large instanton size is required to reproduce the physical axial
coupling.  %\query{why?}From the holographic point of view, this is due to the large quantum
%corrections when quantizing the size as a collective coordinate.
From such an instanton configuration, the predictions for other baryon quantities in the
Skyrme model improve the original results quite a lot. Some of the results
reproduce those by quantizing the full instanton solution, while no fine-tuning of
the mesonic parameters are needed. From these we conclude that the baryons in the
low energy region can be effectively described by large instantons.
%In particular, the instanton size is certainly of physical meaning, not just
%{``}the Cashire cat smile".\query{in bulk, it's point-like on the boundary}

\end{abstract}

 \maketitle

\section{Introduction}
The idea of baryon as a soliton of non-linear interacting mesons, now known as
Skyrmion, was realized by Skyrme long
ago~\cite{Skyrme:1961vq,Skyrme:1961vr,Skyrme:1962vh}. It didn't attract much
attention until Witten showed that baryons are indeed soliton-like objects in the
large-$N_c$ limit~\cite{Witten:1979kh}. However, the phenomenological success of
the Skyrme model and its various extensions is not quite satisfactory (for reviews,
see, e.g.~\cite{Meissner:1985qs,Zahed:1986qz,Meissner:1987ge,Weigel2008}). For
example, in order to obtain the experimental values of the nucleon and $\Delta$
masses, one has to choose an unreasonably small value for the pion decay
constant~\cite{Adkins:1983ya,Jackson:1983bi}.~\footnote{Although the nucleon mass
decreases due to the quantum fluctuations of the pion fields, it is still too large
when the pion decay constant takes its empirical value.~\cite{Zahed:1986va}.}
%Also the axial coupling is much
%smaller than the experimental value. It was argued in \cite{Jackson:1983bi} that
%the axial coupling suffers from large $1/N_c$ corrections.
An interesting
observation was made later by Atiyah and Manton~\cite{Atiyah:1989dq}, who showed
that the holonomy of an instanton along the Euclidean time gave an accurate
approximation to the Skyrmion solution, with the total energy differs by only
$1\%$.

This kind of approach has recently revived due to the advocation of the AdS/CFT
correspondence~\cite{Maldacena1998,Polyakov1998,Witten1998a}. In its original
form~\cite{Maldacena1998}, this correspondence conjectures that the type {II}B
string theory on AdS$_5\times$S$^5$ is dual to the N$=4$ Super Yang--Mills (YM)
theory on the boundary Minkowski spacetime. Though not proven rigorously, this
conjecture has been checked to be valid in many aspects. This encourages people to
find a dual string theory for the realistic QCD in the same spirit. A great success
was made by Karch and Katz who suggested that dynamical fundamental fields could be
introduced by adding flavor branes to the color background~\cite{Katz2002}.
Spontaneous chiral symmetry breaking can then be made manifest as the joining of
the flavor branes with their anti-partners in the deep bulk~\cite{Sakai:2004cn}.
The corresponding Goldstone states are explicitly given by the holonomy of
holographic component of the gauge field on the brane, along the holographic
direction~\cite{Son2004}. The Klein--Kaluza states of the other components give rise
to a tower of vector/axial mesons, while scalar mesons appear from the fluctuations
of the branes in the transverse directions~\cite{Kruczenski:2003be}. Thus the meson
dynamics can be systematically described by the effective action of the flavor
branes in the curved color background. Interestingly, the Skyrme model naturally
arises when one truncates the whole action to the pion modes only, with the
coefficients determined by the 5D metric~\cite{Sakai:2004cn}. On the other side,
the Wess--Zumino--Witten (WZW) term and the anomalous couplings of the mesons
originate from the Chern--Simons (CS) part of the brane
action~\cite{Witten1998a,Sakai:2004cn,Sakai:2005yt}.

Baryons appear in this holographic duality as branes wrapping over some internal
spaces, and become instantons when dissolving into the flavor
brane~\cite{Witten:1998xy}. With the previous identification of the pion states,
the relation between the Skyrmion and the instanton holonomy is then explicitly
established, and the instanton number becomes exactly the winding number of the
Skyrmion irrespective of the background metric~\cite{Son2004}. Thus one could in
principle study baryon properties by studying the instanton configuration on the
curve space. The instanton solutions of the flat space YM action have been studied
extensively in the
literature~\cite{Belavin:1975fg,tHooft:1976fv,Witten:1976ck,Atiyah:1978ri}. In this
case the topological energy bound is saturated by the (anti-)self-dual
instanton solution with an arbitrary size. When a non-zero curvature of the %\query{???} %
space is turned on, the bound can only be attained by the flat instanton solution with an
infinitesimal size~\cite{Hong:2007kx,Hata:2007mb,Pomarol:2007kr}. Recently this
fact was utilized by Sutcliffe to establish a BPS Skyrme model for the
nuclei~\cite{Sutcliffe:2011ig}.%since in the nuclei the binding is very weak.
However, this leads to a rather small value for the classical Skyrmion energy,
and an unreasonably large spin contribution due to the small instanton size.
From the 4D point of view, % \query{why? not clear...}%
this means that inclusion of an infinite tower of vector mesons will push the
instanton to zero size. It was shown that including a single vector meson already
decrease the instanton size by about $30\%$~\cite{Sutcliffe:2010et}.
Correspondingly, the classical total energy and energy radius get reduced by about
$10\%$~\cite{Nawa:2006gv}. Fortunately, the appearance of the CS term naturally
provides a repulsive force which prevents the instanton from
collapsing~\cite{Hata:2007mb}.
It is analogous to the 4D Skyrme model with the $\omega$ field included through
the WZW term~\cite{Adkins:1983nw,Meissner:1986vu,Meissner:1986ka}.
The classical configuration of the instanton is then determined by minimizing the
sum of the YM and the CS energy. In a specific string theory
construction~\cite{Sakai:2004cn}, the YM action was considered to be proportional
to the 't Hooft coupling $\lambda$, and the CS term only perturbates the instanton
to a small size of order $1/\sqrt{\lambda}$~\cite{Hong:2007kx,Hata:2007mb}. In this
case the quantum contribution would be so
 large that one has to fine-tune the parameter to fit the nucleon mass~\cite{Hata:2007mb}, see also ref.~\cite{Sutcliffe:2010et}.
Though the classical instanton size seems to be
suppressed, it attains large quantum corrections when treated as a collective
coordinate~\cite{Hata:2008xc,Hashimoto:2008zw}. This in turn induce large
corrections to the axial coupling and  pushes it close to the observed result,
realizing in some sense the large $N_c$ argument in \cite{Jackson:1983bi} and in
\cite{Hong:2007kx}. Furthermore, phenomenological analysis %\query{not quite correct. \cite{Pomarol:2008aa}: Skyrmion size $\rho\sim\gamma^{2/3}L\ll L$} %
in general backgrounds seems to favor an instanton configuration of a large size, and ensures the calculability of the approach~\cite{Pomarol:2008aa}.
%Numerical methods are then employed to study the complicated solutions.
%The instanton
%configuration can then be analyzed through the 't Hooft coupling
%expansion~\cite{Hashimoto:2008zw}.

In general, it is quite difficult to find the analytic instanton solution in the full YM-CS theory in a curved spacetime.
Based on the 't Hooft coupling expansion, an approximate instanton solution was constructed by combining the
flat-space solution in the infrared and the linearized solution in the ultraviolet~\cite{Hashimoto:2008zw}. However, it was pointed out in ref.~\cite{Cherman:2009gb} that the holographic currents evaluated on this solution fails to reproduce the correct infrared behavior of the
electromagnetic baryon form factors, see also ref.~\cite{Cherman:2011ve}. People also attempted to lift the Noether currents in the Skyrme
model to the 5D theory, but met with problems~\cite{Hata:2008xc}. One then has to employ numerical techniques to obtain the exact solution~\cite{Pomarol:2008aa}. There are also some approximate approaches on this subject, either by expressing the soliton through an effective
Fermion field~\cite{Hong:2007kx}, or by truncating the full theory to the original Skyrme model or its generalized version~\cite{Nawa:2006gv,Sutcliffe:2010et}. In this paper we will try to go a little further in the soliton
description of baryons, combining some of these achievements. We will first show that the infrared
properties of baryons are correctly reproduced in the instanton description,
extending Atiyah and Manton's early observation. In doing so we immediately find
that the axial coupling $g_A$ is proportional to square of the instanton size.
Considering the large quantum corrections to the instanton size in the
holographic approach, we are then led to study the Skyrmion generated from a
large instanton. As a first approximation, we choose to use the flat-space
instanton solution and truncate it to the pion sector. Replacing the original
numerical solution for the Skyrmion by the instanton-generated one, we
recalculated various static properties of the nucleons and Deltas. Quite unexpected,
with the instanton size as the only parameter we obtain a very good description
of most of the quantities. For some observables a better agreement with experimental data
is achieved than those by quantizing the full instanton solution. %\query{what is a full instanton, where and how??} %

The outline of the paper is as follows. In the next section we briefly show the
holographic description of the meson sector, focusing on the framework
implementing nonlinear realization of chiral symmetry. In Sec {III} we review
Atiyah and Manton's observations, and show how these can be explicitly established
in the holographic approach. In Sec.{IV} we show the properties of the Skyrmion
generated from a large instanton. Finally we summarize and outlook for
possible extensions.

\section{Holographic formalism with nonlinear chiral symmetry realization}
The holographic formalism that implement nonlinear chiral symmetry realization in a
holographic way was pioneered in~\cite{Son2004}, and developed later in
\cite{Sakai:2004cn,Sakai:2005yt} and \cite{Hirn:2005nr}. One way to realize the chiral symmetry
breaking is through the boundary conditions in the infrared \cite{Hirn:2005nr}.
From the top-down construction in \cite{Sakai:2004cn,Sakai:2005yt}, one can further attribute
this to the coupling between the flavor branes and the anti-branes in the deep
infrared region.

Following~\cite{Son2004}, we consider the 5D YM action
\begin{equation}\label{SfgF}
  S = -\Tr\int\!dz\,d^4x\,\Bigl(-f^2(z)F_{z\mu}^2
      +\frac1{2g^2(z)} F_{\mu\nu}^2\Bigr),
\end{equation}
where $f(z)$ and $g(z)$ are determined by the metric and the dilaton field.
Without loss of generality, the two boundaries of $z$ can be taken to be
$-\infty$ and $\infty$. They should be considered as the same $4D$ spacetime,
but separated from each other in some extra dimension as in~\cite{Sakai:2004cn}.
To be more specific, we use the {``}$\cosh$" background given by
\begin{equation}\label{SfgF}
ds^2 = - dz^2 + \Lambda^2\cosh^2\Lambda z \eta_{\mu\nu}dx^\mu dx^\nu
\end{equation}
as an example, which can be thought of as two pieces of cut-off anti-de Sitter
spacetime smoothly connected in the infrared. The functions in this background,
after a scaling of $z$ from~\cite{Son2004}, are given by
\begin{eqnarray}
  g^2(z) &=&  g_5^2/\Lambda\,,\\
  f^2(z) &=&  \Lambda \cosh^2 (\Lambda z)/g_5^2.
\end{eqnarray}
To obtain a finite 4D action for the physical modes, the field strength has to
vanish at the two boundaries. One can then choose a gauge configuration in which
the gauge potential also vanishes asymptotically. The residual gauge
transformation $g(x^\mu,z)$ should approach constant boundary values,
$\lim_{z\to\pm\infty}g(x^\mu,z)=g_\pm$, which are then interpreted as elements
of the chiral symmetry group $U(N_f)_{\rm{R}}$ and  $U(N_f)_{\rm{L}}$
respectively. %On the other hand, the physical modes are manifest in the %\query{???} %
The axial gauge $A_z=0$ can be obtained by applying the gauge transformation
with the gauge function given by
\begin{eqnarray}
\xi^{-1}(x^\mu,z)= P \exp\left\{
i\int_{0}^z dz'\, A_z(x^\mu,z')
\right\} \ ,
\label{ginv}
\end{eqnarray}
where $P$ denotes path ordering. Then, the boundary conditions for the gauge potential $A_\mu$ becomes
\begin{eqnarray}
A_\mu(x^\mu,z)\rightarrow i\xi_\pm(x^\mu)\partial_\mu \xi^{-1}_\pm(x^\mu) \ .
~~~(\mbox{at}~z\rightarrow \pm\infty),
\label{asym2}
\end{eqnarray}
with $\xi_\pm(x^\mu)=\xi(x^\mu,\pm\infty)$. Setting $A_z=0$ in the action, the
equation of motion for $A_\mu$ can be derived to be %\query{different from Eq.(6.6)
%in \cite{Son2004}}
\begin{equation}
  \partial_z\left[\cosh^2 (\Lambda z)\, \partial_z A^\mu(q,z)\right] = -q^2 A^\mu(q,z) \,
\end{equation}
in the 4D momentum space. The normalizable (with
vanishing boundary value) solutions are given by
\begin{equation}\label{bncosh}
  \psi_n(z) \propto
  \,\frac{P^1_n(\tanh \Lambda z)}
  {\cosh \Lambda z}\,,
\end{equation}
where $P^1_n$ are the associated Legendre functions. The corresponding eigenvalues
are
\begin{equation}\label{mncosh}
  q^2=m_n^2 = n(n+1)\Lambda^2\, \qquad n=1,\, 2,\ldots \label{eq:mass}
\end{equation}
In accordance with the identification of the chiral symmetry transformation,  one
can choose $(x^i,z)\to (-x^i,-z)$ as the 4D parity transformation. Then the above
solutions have definite parities, and one can see that the odd $n$ states
correspond to the vector mesons and  even $n$  to axial mesons~\cite{Son2004}.
In addition to these, there are also two non-normalizable solutions at $q^2=0$, namely
$1$ and $\tanh(\Lambda z)\equiv\psi_0(z)$, that also give rise to physical
configurations. %\query{What kind of physical configurations?}
Summing all these up,
the gauge potential with the above boundary conditions can be decomposed
as~\cite{Sakai:2004cn,Sakai:2005yt}%{\bf ?}%\query{definition of $B_\mu$?}
\begin{eqnarray}
A_\mu(x^\mu,z)=i \xi_+(x^\mu)\partial_\mu \xi_+^{-1}(x^\mu)\psi_+(z)
+i \xi_-(x^\mu)\partial_\mu \xi_-^{-1}(x^\mu)\psi_-(z)
+\sum_{n\ge 1} B_\mu^{(n)}(x^\mu) \psi_n(z) \ ,
\label{nonAexp0}
\end{eqnarray}
where
\begin{equation}
\psi_\pm(z)\equiv \frac{1}{2}(1\pm \psi_0(z)) \ .
\end{equation}
From this one then obtains the effective action of the pion and all the vector/axial
mesons, as shown in~\cite{Sakai:2004cn}. Here we omit all the vector fields
$B_\mu^{(n)}$ and focus on the pion freedom only. The pion field can be defined
as~\cite{Son2004}
\begin{eqnarray}
e^{2i\pi(x^\mu)/f_\pi}\equiv U(x^\mu)\equiv \xi_- \xi^{-1}_+ =P \exp\left\{
i\int_{-\infty}^{\infty} dz'\, A_z(x^\mu,z')
\right\} \ ,
\label{eq-U}
\end{eqnarray}
which transforms as $U(x^\mu)\rightarrow g_+ U(x^\mu) g_-^{-1}$ as expected. By
using the residue gauge symmetry one can further choose $\xi_-(x^\mu)=1$, then we
have~\cite{Sakai:2004cn}
\begin{eqnarray}
A_\mu(x^\mu,z)=i U^{-1}(x^\mu)\partial_\mu U(x^\mu)\psi_+(z)\ .
\label{nonAexp}
\end{eqnarray}
Substituting this into the action we arrive at the Skyrme
model~\cite{Skyrme:1961vq}%\query{CHECK}
\begin{eqnarray}
S=\int d^4 x\left(-\frac{f_\pi^2}{4}\tr\left(
 U^{-1}\partial_\mu U\right)^2+\frac{1}{32 e^2}
\tr\left[U^{-1}\partial_\mu U,U^{-1}\partial_\nu U\right]^2
\right) \ ,
\label{eq-Skyrme}
\end{eqnarray}
with the parameters given by
\begin{eqnarray}
f_\pi^2&=& 4\int \mathrm{d}z f^2(z) \, \partial_z^2 \psi_+(z) =\frac{2 \Lambda^2}{g_5^2}, \\
e^{-2}&=& \int \mathrm{d}z g^{-2}(z) (1-\psi_0^2(z))^2=\frac{4}{3g_5^2},
\end{eqnarray}
where the first equation was already given in Ref.~\cite{Son2004}. One can further
show that the WZW term of the pion and the various anomalous couplings of the
mesons can be derived from the CS term on the 5D
side~\cite{Sakai:2004cn}.

Now let us show the phenomenological predictions in this model. There are only two
unknown parameters, $\Lambda$ and $g_5$. From the experimental $\rho$-meson mass
one can fix $\Lambda$ as $\Lambda=0.55 ~\rm{GeV}$. Matching the asymptotical
behavior of the vector correlator to QCD calculation one obtains
\begin{equation}
g^{-2}_5=\frac{N_c}{24\pi^2}.
\end{equation}
With these values the predictions for some other quantities, including the
low-energy constants $L_1^r,L_2^r$ and $L_3^r$ in ${\cal O}( p^4)$ chiral
perturbation theory (CHPT) and the mass of the axial meson $a_1$, are displayed in
Tab.\ref{tab-meson}.
%----------------------------------------------------
\begin{table}[t!]
\centering
\begin{tabular}{|c|c|c|c|c|c|c|}
\hline
& $f_\pi$ (GeV) & $~~~e~~~$  &  $~L_1^r(10^{-3})~$ &
 $~L_2^r(10^{-3})~$ & $~L_3^r(10^{-3})~$ & $~m_{a_1}$(GeV) \\ \hline
{``}cosh"   &  $0.087$  & $7.70$ & $0.53$ & $1.1$ & $-3.2$ & $1.335$ \\
Empirical  &  $0.0922$ &  ---    & $0.43\pm0.12$ & $0.73\pm0.12$ & $-2.35\pm0.37$ & $1.23\pm 0.04 $\\
\hline
\end{tabular}
\caption{Predictions of the {``}cosh" model for the mesonic observables, in
comparison with the empirical values given in the last row. The pion decay constant
and $a_1$ mass are from Ref.~\cite{Nakamura:2010zzi}, and the values of $L_i^r$'s
at the scale $\mu=m_\rho$ are taken from Ref.~\cite{Amoros:2001cp}.} %bi95
 \label{tab-meson}
 \end{table}
%----------------------------------------------------
The accuracy is typical in this kind of formalism, since the dependence of the
results on the backgrounds is very weak as shown in~\cite{Becciolini:2009fu}. In
fact, since the quark mass terms are not included here, it is more appropriate to
compare the result of $f_\pi$  with that in the chiral limit, which is $87$ MeV
derived from ${\cal O}( p^4)$ CHPT~\cite{Gasser:1983yg}. The values of
$L_{1,2,3}^r$ taken from Ref.~\cite{Amoros:2001cp} were obtained from a
$\mathcal{O}(p^6)$ fit, and their $\mathcal{O}(p^4)$ values are
$0.38\times10^{-3}$, $1.59\times10^{-3}$ and $-2.91\times10^{-3}$, respectively.
The $L_i^r$'s, absorbing the one-loop divergences in CHPT, are scale-dependent. As
usually done in the resonance saturation model of the low-energy
constants~\cite{egpr}, here we assume the predicted values should be compared with
those at $\mu=m_\rho$ since $\Lambda$ is fixed through the $\rho$-meson mass.
Notice that as shown in Ref.~\cite{egpr}, $L_3^r$ also contains a small
contribution from the scalar meson exchange, $L_{3S}^r=0.6\times10^{-3}$, which is
not included in our approach.
%, since some of the latter still suffer from large uncertainties.
%\query{refs. for the exp values...}
We will use the values of $f_\pi$ and $e$ obtained here in the following sections
rather than the experimental values for consistency.

\section{From Skyrmions to holographic instantons}
With great intuition, in the 1960s Skyrme proposed to use the nonlinear action
(\ref{eq-Skyrme}) to describe baryons, treating them as collective excitations of
the Goldstone bosons. The topological number of the meson configuration is
identified with baryon number, and he conjectured that the fluctuations should be
quantized as fermions. All these have been confirmed by later
investigations~\cite{Finkelstein:1968hy,Witten:1979kh,Balachandran:1982dw,Witten:1983tx}.

Now let us take a careful look at some dynamical properties of the model.
For the static configuration, the energy functional is given by
\begin{eqnarray}
E_{\rm{cl}}=-\int d^3 x\left(\frac{f_\pi^2}{4}\tr[L_i^2]+\frac{1}{32 e^2}
\tr\left[L_i,L_j\right]^2
\right) \ ,
\end{eqnarray}
where the left currents $L_\mu=U^{-1}\partial_\mu U$ have been used. This
functional was found to be bound from below as
%\query{$\frac{6\pi^2}{7.7}f_\pi\sim2.5\pi f_\pi\sim 700$~MeV}
\begin{equation}
E_{\rm{cl}} \geq 6 \pi^2 f_\pi |B|/e,
\end{equation}
where $B$ is the topological charge related to the baryon current
\begin{equation}
B^\mu=\frac{i\varepsilon^{\mu\nu\alpha\beta}}{24\pi^2}\tr[L_\nu L_\alpha L_\beta].\label{eq:Bmu}
\end{equation}
Due to the metrical difference of the space and the group
manifold~\cite{Manton:1986pz}, the bound can not be saturated using the
(anti)self-dual solutions. One then has to solve the equation of motion directly,
which reads
\begin{equation}
\partial^\mu L_\mu- \frac{1}{4e^2f_\pi^2}\partial^\mu [L_\nu,[L_\mu,L_\nu]]=0. \label{eq-Seq}
\end{equation}
Due to the highly non-linearity of this equation, Skyrme suggested to focus on the so-called hedgehog configuration
\begin{equation}
U(\vec{x})=\exp(i\vec{\tau}\cdot \hat{r} F(r))=\cos F(r)+i\vec{\tau}\cdot \hat{r} \sin F(r).
\end{equation}
Then the equation of motion simplifies into a radial equation of $F(r)$
\begin{equation}
F^{''}+\frac{2}{r}F^{'}-\frac{\sin 2F}{r^2}-\frac{1}{e^2f_\pi^2}\left[\frac{\sin 2F \sin ^2F}{r^4}-\frac{{F^{'}}^2\sin 2F}{r^2}
-\frac{2F^{''}\sin^2F}{r^2}\right]=0.
\end{equation}
In the $B=1$ sector, combining the above equation and the topological constraint
one usually choose the following asymptotic behavior for the chiral angle $F(r)$:
\begin{eqnarray}
F(r) &\to& \pi-\alpha r, ~~r\to 0, \nonumber\\
F(r) &\to& \beta/r^2, ~~r\to \infty. \label{eq-Fbc}
\end{eqnarray}
The numerical calculations were first undertaken by Skyrme himself and further
developed by many others, see e.g.
Refs.~\cite{Adkins:1983ya,Jackson:1983bi,Meissner:1983}. The minimum energy is
about $E_{\rm{cl}}=72.9 f_\pi/e$, $23\%$ above the topological bound. It worths
pointing out that a fine-tuned solution with the above asymptotic behavior gives a
quite close result~\cite{Manton:1986pz}.

Skyrme's work did not receive much attention until it was realized that in the
$1/N_c$ expansion baryon should really be interpreted as solitons in the meson
theory~\cite{Witten:1979kh}. While unable to sum up the planar amplitudes, the
Skyrme model is the natural action to start with. The quantization of the model was
systematically developed using the method of collective coordinates or moduli
quantization, and various properties of the ground state baryons were investigated
\cite{Adkins:1983ya}. To do this, one considers the rigid rotation $U=A(t) U_0
A^{-1}(t)$ of the classical hedgehog solution $U_0=\exp(i\vec{\tau}\cdot \hat{r}
F(r))$, and treats $A$ as collective coordinates. Parameterizing $A$ as $A=a_0+i
\vec{a}\cdot \vec{\tau}$ with $a_0^2+\vec{a}^2=1$, the Lagrangian then becomes
\begin{equation}
L=-E_{\rm{cl}}+2I_0 \sum^3_{i=0}(\dot{a}_i)^2,
\end{equation}
with $I_0$ a constant determined by $F(r)$. Choosing the suitable quantization
framework, one obtains a series of Fermion states of quantum numbers $I=J$, with
$I$ and $J$ being isospin and spin respectively. The
states with $I=J=1/2$ and $I=J=3/2$ are then identified with the nucleons and
Deltas, with masses
\begin{equation}
M_N=E_{\rm{cl}}+\frac{1}{2I_0}\frac{3}{4},\,\,M_\Delta=E_{\rm{cl}}+\frac{1}{2I_0}\frac{15}{4}.
\end{equation}
The Noether currents of the model, expressed in terms of the chiral angle and the
collective coordinates, are as follows~\cite{Adkins:1983ya,Zahed:1986qz,Weigel2008}
\begin{eqnarray}
B_0&=&b(r)\,\,,B_i=b(r)\varepsilon_{ijk}\Omega_jx_k,\nonumber\\
V_0^a&=&- v(r)D_{ai}\Omega_i\,\,,V_i^a=\frac{v(r)}{r^2}\varepsilon_{ijk}x_jD_{ak},\nonumber\\
A_i^a&=&\left[a_1(r)\delta_{ik}+a_2(r)\hat{x}_i\hat{x}_k\right]D_{ak}  \label{eq-current}
\end{eqnarray}
where $\vec{\Omega}\equiv -i\tr[A^{-1}\dot{A}(t)\vec{\tau}]$, $D_{ab}(A)\equiv \tr[\tau_aA\tau_bA^{-1}]/2$, and
\begin{eqnarray}
b(r)&=&-\frac{1}{2\pi^2}\frac{\sin ^2F}{r^2}F^{'}\nonumber\\
v(r)&=&\sin^2F\left[f_\pi^2+\frac{1}{e^2}\left({F^{'}}^2+\frac{\sin^2F}{r^2}\right)\right]\nonumber\\
a_1(r)&=&\frac{\sin 2F}{2r}\left[f_\pi^2+\frac{1}{e^2}\left({F^{'}}^2+\frac{\sin^2F}{r^2}\right)\right]\nonumber\\
a_2(r)&=&-a_1(r)+F^{'}\left[f_\pi^2+\frac{2}{e^2}\frac{\sin^2F}{r^2}\right]
\end{eqnarray}
Sandwiching these between the baryon states, one can study various properties of
them. For example, the axial coupling can be obtained by direct evaluating the
expectation value of the integral $\int d^3 x A_i^a(x)$ on the nucleon state.
Alternatively, one can transform this into a surface integral at infinity using
the conservation equation, then the axial coupling is directly related to the
asymptotic behavior of $F(r)\to \beta/r^2$ when $r\to \infty$ as~\cite{Zahed:1986qz}
\begin{equation}
g_A=\frac{8\pi}{3}f_\pi^2\beta.\label{eq-gA}
\end{equation}
By comparing this behavior with that of the pion in the old fashioned pion-nucleon
Lagrangian one can further deduce the Goldberger-Treiman relation. Unfortunately,
the phenomenological results are not quite good. Generally, the parameters fitted
from the baryonic sector are inconsistent with those in the mesonic sector and the
experimental values, and the prediction of $g_A$ is normally small. %In
%\cite{Jackson:1983bi} it was argued that this prediction corresponds to $N_c\to
%\infty$, and that for $N_c=3$ should be enhanced by $(N_c+2)/N_c$.

The relation between the Skyrmion configuration in $\mathbb{R}^3$ and the instanton
solution in $\mathbb{R}^4$ was first found by Atiyah and Manton in
1989~\cite{Atiyah:1989dq}. They identified the holonomy of an instanton along the
time direction as a Skyrmion configuration,
\begin{equation}
U(\vec{x})=T\exp\left(-\int_{-\infty}^\infty A_t(\vec{x},t)dt\right), \label{eq-holonomy}
\end{equation}
where $T$ denotes time ordering. Employing the 't Hooft instanton
construction~\cite{tHooft:1976fv}, this gives
\begin{equation}
F(r)=\pi[1-(1+\rho^2/r^2)^{-1/2}].\label{eq-CA0}
\end{equation}
The minimum energy from this profile occurs at $\rho^2=2.11/(e^2f_\pi^2)$,  and
is only $1\%$ above the numerical value. Notice also that this function has the
right asymptotic behavior as required.

However, the dynamical reason for the
relation (\ref{eq-holonomy}) was then not clear. Only after the discovery
of the AdS/CFT correspondence and the development of the framework of holographic QCD,
this relation was explicitly established~\cite{Son2004,Sakai:2004cn}. Based on the
large$-N$ expansion and the AdS/CFT correspondence, Witten proposed that baryons
should by described as branes wrapping on some internal manifolds. Effectively,
these branes can be viewed as flavor instantons from the 4D spacetime point of
view. From the discussion in the previous section, this instanton becomes a
Skyrmion when we truncate it to the Goldstone sector. Also the holonomy should be
taken in the holographic direction, rather than the time coordinate. Explicitly,
the Skyrmion profile can be obtained as Eq.(\ref{eq-U})
\begin{eqnarray}
U(\vec{x}) =P \exp\left\{
i\int_{-\infty}^{\infty} dz'\, A_z(\vec{x},z')
\right\} \ .
\label{eq-SkIn}
\end{eqnarray}
A direct check of this relation is given by the
equality of the Skyrmion winding number and the instanton number, shown
in~\cite{Son2004}.

\section{Skyrmion properties from holographic instanton}
%Since the proposal of the holographic instanton
%picture~\cite{Son2004}, there have been extensive studies of the
%baryonic sector. Due to the complexity of the instanton equation on curved space,
%only limited progresses have been made. To deal with the technical problems, there
%are mainly two kinds of approaches up to now. The first is the 't Hooft coupling
%expansion, valid only in some specific construction~\cite{Hashimoto:2008zw}.  One
%start from the flat-space instanton solution, and include the curvature effects
%order by order in the expansion. The other approach employs numerical technology,
%and does not rely on the flat-space approximation~\cite{Pomarol:2008aa}. To test
%these approaches, Cohen et. al proposed some kind of model-independent test
%concerning the large distance behavior of the baryon form factors in the chiral
%limit~\cite{Cherman:2009gb}. A large $N_c$ relation of baryon form factors at
%large distance was derived using %the
%Skyrmion property (\ref{eq-Fbc})
%chiral symmetry of QCD. %and the structure of the currents (\ref{eq-current}).
%Exact the same power behavior for the form factors was found from the asymptotic
%instanton solutions in~\cite{Pomarol:2008aa}. However, the corresponding form
%factors in the Sakai-Sugimoto construction generally exhibit exponential behavior
%in the infrared, contrary to the expected power form. Based on this fact, the
%reliability of the holographic description of baryons in terms of intantons was
%doubted~\cite{Cherman:2009gb}.

In this section we will attempt to improve the Skyrme model from the holographic
approach. Recently, using large $N_c$ chiral perturbation theory, Cherman et al.
found a model-independent relation concerning the large distance behavior of the
baryon form factors in the large $N_c$ and chiral limit~\cite{Cherman:2009gb}.
These relations are shown to be valid in the Skyrme model. So first
 we will show explicitly that these relations are valid in the holographic approach.
There are two ways to obtain the infrared limit of baryon form factors in the full theory. One
may first study them by using the full intanton solution as in \cite{Hashimoto:2008zw}
and \cite{Pomarol:2008aa}, and then take the IR limit~\cite{Cherman:2009gb,Cherman:2011ve}.
Alternatively, one can just start from the flat-space instanton solution,
and truncate it to the Goldstone sector to get an approximate Skyrmion solution. The IR properties of those form factors can then be
obtained in the same way as in the Skyrme model~\cite{Adkins:1983ya}. These two approaches should give the same IR results.
In the previous section we have shown that the holographic action reduces to the
Skyrme model when truncated to the massless sector, thus the current structure at
large distance should always be described as in (\ref{eq-current}). To ensure the
form factor relations in~\cite{Cherman:2009gb}, we only need to show that the
Skyrmion configuration generated from the flat-space instanton has the correct IR
behavior. As shown in the previous section, this has been found to be true long ago
by Atiyah and Manton. So in general the form factor relations in
\cite{Cherman:2009gb} will be valid in the instanton description. One may ask why the currents derived in ref.~\cite{Hashimoto:2008zw} fail to produce these relations. It was pointed out in~\cite{Cherman:2009gb} that this may be due
to the strongly suppression of the pion loop contributions in the 't Hooft
coupling expansion. Indeed, since the pion loops are discarded, the isovector mean
square radius of the nucleon is determined by the vector fluctuations and remains finite in the
chiral limit~\cite{Hashimoto:2008zw}. It was also shown in \cite{Hata:2008xc} that
the holographic currents constructed using the 't Hooft expansion do not recover
those in the Skyrme model in the infrared. Alternatively, one may attempt to
lift the Skyrmion currents (\ref{eq-current}) to the five-dimension theory in the same spirit as for the
baryonic current~\cite{Son2004}, but both uniqueness and gauge invariance
are not guaranteed~\cite{Hata:2008xc}. Actually there is a direct way to verify our statement that the instanton description gives the correct IR behavior of  baryon form factors. One can do a large $r$ expansion of the instanton equations and find the solutions order by order~\cite{Panico:2008it}. Substituting the leading order result to the holographic currents gives exactly the correct form factor relations, which has been proved in the so-called hard-wall model~\cite{Panico:2008it,Cherman:2009gb}, and recently for the Sakai-Sugimoto model~\cite{Cherman:2011ve}.

%It seems that only numerical analysis
%succeeds in doing this~\cite{Pomarol:2008aa}.

Now let us turn to the issue of the instanton size. As discussed extensively in
\cite{Hong:2007kx,Hata:2007mb} and \cite{Pomarol:2007kr}, the SU$(2)$ YM instanton
in a curved space will generally collapse to zero size, in contrast to the
situation in the flat space. This can be cured when a nontrivial CS term is
included, which can be realized by enlarging the gauge group to U$(2)$. In this
case, the induced coupling between the Skyrmion and the isoscalar current provides
a Coulomb-like repulsive force which prevents the Skyrmion to shrink. This is very
similar to the early attempt to stabilize the Skyrmion by introducing the $\omega$
meson~\cite{Adkins:1983nw,Meissner:1986js}. The actual size will be determined by
the competition of the curvature effect and the Coulomb force. This requires the
minimization of the energy functional of the full YM-CS system. Some estimates of
the instanton size were made in \cite{Hong:2007kx,Hata:2007mb} and
\cite{Pomarol:2007kr}. In the Sakai-Sugimoto model, the YM action dominates the CS
term by an order of the 't Hooft coupling $\lambda$, and the instanton size is
suppressed by $1/\sqrt{\lambda}$ compared to the typical size of the model. A
manifestation of the small instanton size is the large value for the mass
splitting~\cite{Hashimoto:2008zw}. However, when quantized as one of the collective
coordinates, the size acquires large quantum collections, as large as $60\%$ even
for the nucleons~\cite{Hata:2008xc}. In general backgrounds the instanton size is
argued to be large, much larger than inverse of the cut off scale, which guarantees
the calculability of the instanton configuration~\cite{Pomarol:2007kr}. There are also arguments that the
instanton size is not physical and can be {``}gauged away", just as the bag radius
in the bag model~\cite{Nielsen:2009pw}.

Since the full analytical instanton solution in the YM-CS system is still
unaccessible right now, we will attempt to determine the instanton size from
phenomenological considerations. We will focus on quantities which are
sensitive to the massless sector only, so we can use the flat-space solution
safely. It turns out that the axial coupling $g_A$ is a quite good candidate.
However, some ambiguities should be clarified here. Since the instanton-generated
%\query{too many ``since"...}
Skyrmion actually does not satisfy the equation
(\ref{eq-Seq}) exactly, the axial current induced from it will not be conserved in general.
Thus the two methods used to calculate $g_A$ may not be equivalent any more.
The direct calculation from the current expectation value involves the sub-leading terms
in the Skyrmion solution, which may not be well approximated from
the instanton profile. So we choose to estimate
$g_A$ from the asymptotic behavior of the current. In doing so the
Goldberger-Treiman relation will also be guaranteed. A direct calculation of the
holonomy (\ref{eq-SkIn}) from the BPST instanton \cite{Belavin:1975fg}
gives~\cite{Sutcliffe:2010et,Nielsen:2009pw}
\begin{equation}
F(r)=-\pi(1+\rho^2/r^2)^{-1/2},\label{eq-CA1}
\end{equation}
which differs from (\ref{eq-CA0}) by a constant shift of $\pi$. This corresponds to
a different choice of the boundary values from (\ref{eq-Fbc}) and does not affect
the results. Taking the large-$r$ expansion, one obtains
\begin{equation}
\beta=\pi\rho^2/2.
\end{equation}
Combing this with (\ref{eq-gA}) gives the relation between the instanton size and
the axial coupling
\begin{equation}
g_A=\frac{4\pi^2}{3}f_\pi^2\rho^2.
\end{equation}
This is exactly the result obtained by quantizing the full instanton solution in
\cite{Hata:2008xc,Hashimoto:2008zw}. From the experimental value $g_A=1.27$~\cite{Nakamura:2010zzi} and the
parameters given in Tab. \ref{tab-meson}, one fixes the size
\begin{equation}
\tilde{\rho}^2\equiv\rho^2\,(e^2f_\pi^2)=5.72.\label{eq-instanton3}
\end{equation}
Equivalently, we have $\rho\approx 0.7 ~\mbox{fm}$, to be compared with the
averaged size $\rho_N=0.67~\mbox{fm}$ on the nucleon state from the quantization procedure~\cite{Hata:2008xc}.
It is almost three times larger than that obtained from minimizing the
Skyrmion energy, where $\tilde{\rho}^2=2.11$, while a rough estimate of the
classical size as in \cite{Hata:2007mb} gives $\tilde{\rho}^2\approx 4.0$. To
confirm the above result, let us take a closer look at it by exploring its
consequences in the Skyrme model. Following the standard procedure
in~\cite{Adkins:1983ya}, the static properties of the nucleon and Delta can be
obtained. One should keep in mind that in doing so we are actually attributing the
same size to the nucleon and Delta. Fortunately the difference between them is of
${\cal O}(1/N_c^2)$~\cite{Hashimoto:2008zw}, and can be neglected. We list the
results for all these quantities in Tab. \ref{tab-baryon} from the instanton
approximation with these two values of $\tilde{\rho}^2$, in comparison with the
experimental values. We have also listed the results in the original Skyrme model~\cite{Adkins:1983ya},
reexpressed with the parameters shown in Tab. \ref{tab-meson}. Note that the
coupling $g_{\pi NN}$ is derived from the Goldberger-Treiman relation, and in
calculating of $g_{\pi N \Delta}$ and $\mu_{N \Delta}$ we have employed the
model-independent large $N_c$ relations~\cite{Adkins:1983ya}
\begin{eqnarray}
g_{\pi N \Delta}&=&\frac{3}{2}g_{\pi NN} \nonumber\\
\mu_{N\Delta}&=&\sqrt{\frac{1}{2}}(\mu_p-\mu_n).
\end{eqnarray}

\begin{table}[t!]
\centering
\begin{tabular}{|c|c|c|c|c|c|c|c|c|c|c|c|}
\hline
 Quantity& $~~E_{\rm{cl}}~~$ & $~~g_A~~$ &  $~~M_N~~$ &
 $~~M_\Delta~~$  & $g_{\pi NN}$ & $\langle r^2\rangle^{1/2}_{I=0}$ & $\langle r^2\rangle^{1/2}_{M,I=0}$ & $~~\mu_p~~$ & $~~\mu_n~~$   &  $g_{\pi N \Delta}$ & $\mu_{N\Delta}$  \\ \hline
Skyrmion         &  $0.83$ &  $0.31$  & $1.11$  & $2.23$ & $3.88$ & $0.31$ & $0.48$ & $0.58$ & $-0.41$   & $5.82$ & $0.70$  \\
Instanton-1      &  $0.84$ &  $0.47$  & $1.05$  & $1.89$ & $5.62$ & $0.32$ & $0.57$ & $0.87$ & $-0.36$   & $8.43$ & $0.87$  \\
Instanton-2      &  $0.94$ &  input   & $1.00$ & $1.23$ & $14.5$ & $0.52$ & $0.94$ & $2.32$ & $-1.96$   & $21.8$ & $3.0$  \\
Experiment       &  ---    &  $1.27$   & $0.939$ & $1.23$ & $13.5$ & $0.81$ & $0.84$ & $2.79$ & $-1.91$  & $20.3$ & $3.3$  \\
\hline
\end{tabular}
\caption{Predictions of baryon properties from the instanton approximation, in
comparison with the experimental
values~\cite{Nakamura:2010zzi,Belushkin:2006qa,deSwart:1997ep} and the original
Skyrmion results~\cite{Adkins:1983ya}, reexpressed with the same parameters.
{``}Instanton-1" stands for Atiyah and Manton's small instanton configuration with
$\tilde{\rho}^2=2.11$, while {``}Instanton-2" the large instanton configuration
with $\tilde{\rho}^2=5.72$. Energy and masses are given in GeV, and the radii in
fm. }
 \label{tab-baryon}
\end{table}

From Tab. \ref{tab-baryon} we see that, when reexpressed in the physical %acceptable
parameters, the original Skyrmion results deviate much from experimental values.
The masses are too large, while all the other physical quantities are too small.
The smallness of these predictions has been pointed out in the original paper, but
the deviation is not so serious since the parameters there have been finely tuned.
The results from the small instanton configuration show a very similar pattern.
Amazingly, when the instanton size increases, all these quantities are pushed
towards the experimental results: the masses are lowered, and all the other
quantities increase. The decreasing of the masses can be easily understood from the
fact that when the instanton grows, the spinning contribution is greatly
suppressed~\cite{Sutcliffe:2010et}. For the specific value $\tilde{\rho}^2=5.72$,
the predictions agree with the experimental values so well that, except for
$\langle r^2\rangle^{1/2}_{I=0}$, all the other predictions deviate less than
$17\%$ from the experimental values. Actually, our results for the radii can be more compactly expressed as
\begin{equation}
\langle r^2\rangle^{1/2}_{I=0}=\sqrt{0.56~\rho^2}, \quad \langle r^2\rangle^{1/2}_{M,I=0}=\sqrt{1.81~\rho^2}.
\end{equation}
While the prediction for the isoscalar magnetic radius is reasonably consistent with the experimental value, the result for the
isoscalar charge radius is somewhat small. This deviation also induces a
large deviation for the isoscalar $g$ factor, $g_{I=0}$, since we have in the
Skyrme model the relation~\cite{Adkins:1983ya}
\begin{equation}
g_{I=0}=\frac{4}{9}\langle r^2\rangle_{I=0}M_N(M_\Delta-M_N).
\end{equation}
Our prediction is $g_{I=0}=0.73$, while the experimental result is $1.76$, instead.
On the other hand, our prediction for the isovector $g$ factor, $g_{I=1}=8.56$, is in
reasonable agreement with experimental value $g_{I=1}=9.41$, confirming the
observation that $g_{I=1}$ is proportional to the square of the average instanton
size~\cite{Hata:2008xc,Hashimoto:2008zw}.

The isoscalar charge radius depends much on the definition of the baryon current and
shows how the baryon charge is distributed.
In fact, it was found that modifying the currents by coupling the $\omega$ meson to
$B^\mu$ in the Skyrme model, the isoscalar radius is enlarged with a change
$\delta\langle r^2\rangle_{I=0} = 6/m_\omega^2$~\cite{Adkins:1983nw,Meissner:1986js}. %and $\delta\langle r^2\rangle_{M,I=0} = 10/m_\omega^2$~\cite{Adkins:1983nw}.
If adding this change naively to the result presented in Table~\ref{tab-baryon}, one would get $\langle
r^2\rangle_{I=0}^{1/2}=0.81$~fm, quite close to the empirical value. %and $\langle r^2\rangle_{M,I=0}^{1/2}=1.23$~fm. The
%former becomes close to the empirical value, while the latter is too large.
%However, recall that we have truncated the model to only keeping the Goldstone
%boson mode, such a modification due to the $\omega$ meson is not self-consistent.
%The inclusion of the vector mesons is relegated to future work.
In ref.~\cite{Meissner:1986js} it was pointed out that this significant change is due to the fact that the
isoscalar photon only couples to the baryon charge indirectly through the $\omega$ field. In other words,
the baryonic current can only be correctly introduced through gauging the $U(1)_V$ symmetry. In the holographic
framework, the baryonic charge density is given by~\cite{Son2004,Hashimoto:2008zw}
\begin{equation}
B^0=\frac{1}{32\pi^2}\int\!dz\,\,
  \epsilon^{\mu\nu\lambda\rho}\Tr F_{\nu\lambda} F_{\rho\sigma}.
\end{equation}
Although this gives the same baryon number as the baryon current (\ref{eq:Bmu})~\cite{Son2004}, the charge distribution is different.
Substituting the BPST solution into this expression, one finds that the isoscalar charge radius
is related to the instanton size as~\cite{Hata:2008xc}
\begin{equation}
\langle r^2\rangle^{1/2}_{I=0}=\sqrt{\frac{3}{2}\,\rho^2}.
\end{equation}
With our estimate (\ref{eq-instanton3}) this gives $\langle
r^2\rangle^{1/2}_{I=0}=0.86~\mbox{fm}$, again close to the experimental result.
However, in the flat-space approximation all the vector/axial fluctuations would be
massless, as con be seen from Eq.~(\ref{eq:mass}) by setting $\Lambda=0$. They also exhibit power
behavior in the infrared and may even dominate over the pion. Indeed one finds that
the so-derived currents  exhibit different IR behavior
from those in Eqs.~(\ref{eq-current}). The divergence of the isoscalar magnetic radius manifestly reflects this fact~\cite{Hata:2008xc}. The currents in
ref.~\cite{Hashimoto:2008zw} suffer from the same problem~\cite{Cherman:2009gb}, which exhibit exponential rather than power behavior in
the IR. So far it is still not clear how to obtain the analytic results of the currents with the
correct IR behavior. As a first approximation, one may follow the procedure in Refs~\cite{Adkins:1983nw,Meissner:1986js} to
include the $\omega$ meson first, and introduce the currents accordingly. We will systematically study the effect of the $\omega$ meson
from the holographic view in a forthcoming paper.
%Thus it is not clear whether such relations will hold when the currents are
%obtained with the correct IR behavior.

\section{Discussion}
In this paper we illustrate how the holographic instanton description of baryons
naturally implements the traditional Skyrmion picture, and correspondingly the
correct infrared behavior of the baryon form factors. In our study, we didn't minimize the energy functional with
respect to the instanton size. Instead, we determine it from the phenomenological analysis,
and found that the instanton should possess a large size. This is in
accordance with the instanton quantization procedure which indicates that the instanton size
suffers from large quantum corrections. Therefore the present treatment could be
considered as a classical approximation to collective coordinate motion~\cite{Manton:2011mi}.
The Skyrmion solution generated from such an instanton configuration gives very good predictions for most of the static
baryon quantities.
%
%One might worry that the nucleons obtained in such a way are not stable. However,
%the instanton by itself is already a solution that satisfies the Bogomolny bound in
%4D Euclidean space. Hence, it is not a priori necessary to request the instanton
%size minimize the energy functional.
%
Thus our approach could serve as a good toy model for various further
applications. One could generalize the present calculation to the study of various
dynamic quantities, as in \cite{Hashimoto:2008zw} and \cite{Panico:2008it}. One
could also extend the present analysis to the case with nonzero quark mass, and
further enlarge the flavor group to SU$(3)$. The including of quark mass in such a
framework has been extensively discussed in the literature, and has been recently
discussed in \cite{Domenech:2010aq}. An interesting idea towards this was given in
\cite{Atiyah:2004nh}, where a Skyrmion with massive pions was approximated by the
holonomy of a Yang--Mills instanton along circles. This seems feasible since in this
approximation the massless limit can be taken smoothly and reproduce the structure
in the chiral limit. One can also improve the present calculation by including the
massive fluctuations, which means that we have to go beyond the flat-space
approximation. We postpone all these for future studies.

\hspace{1cm}

{\bf Acknowledgments}: We would like to thank Pietro Colangelo and
Ulf-G.~Mei{\ss}ner for useful discussions and comments. This work supported in part
by Natural Science Foundation of China under Grant No.~10975144, No.~10735080, by
DFG through funds provided to the SFB/TR 16 and the EU I3HP ``Study of Strongly
Interacting Matter'' under the Seventh Framework Program of the EU.

%\bibliography{Mybib}
%\nocite{*}
\newpage

\end{document}